\documentclass{aa} 
\usepackage{txfonts}
\usepackage{graphicx}
\usepackage{natbib} 
\begin{document} 

\title{VdBH~222: a starburst cluster in the inner Milky Way\thanks{Partially based on observations collected at the European Organisation for Astronomical Research in the Southern Hemisphere, Chile (ESO 089.D-0332, ESO 081.D-0812) and the Australian Astronomical Observatory}\thanks{Tables 4, 5, 7 and 8 are only available in electronic form at the CDS via anonymous ftp to cdsarc.u-strasbg.fr (130.79.128.5) or via http://cdsweb.u-strasbg.fr/cgi-bin/qcat?J/A+A/}}


   \author{A. Marco
          \inst{1}
          \and 
          I. Negueruela
          \inst{1}
          \and
          C. Gonz\'alez-Fern\'andez
          \inst{1,2}
          \and
          J. Ma\'{\i}z Apell\'aniz
          \inst{3}
          \and
          R. Dorda
          \inst{1}
          \and
          J.S. Clark
          \inst{4}
           }

  \institute{Departamento de F\'{i}sica, Ingenier\'{i}a de Sistemas y Teor\'{i}a de la Se\~{n}al. Escuela Polit\'ecnica Superior. University of Alicante. Apdo.99 E-03080. Alicante, Spain\\
              \email{amparo.marco@ua.es}
              \and
              Institute of Astronomy, University of Cambridge, Madingley Road, Cambridge CB3 0HA, United Kingdom
              \and
              Instituto de Astrof\'{\i}sica de Andaluc\'{\i}a-CSIC, Glorieta de la Astronom\'{\i}a s/n, 18008, Granada, Spain
              \and
              Department of Physics and Astronomy, The Open University, Walton Hall, Milton Keynes, MK7 6AA, United Kingdom                    }
   
\date{}
\titlerunning{The starburst cluster VdBH~222}

 
\abstract
   {It has been suggested that the compact open cluster VdBH~222 is a young massive distant object.}
   {We set out to characterise VdBH~222 using a comprehensive set of multi-wavelength observations.}
   {We obtained multi-band optical ($UBVR$) and near-infrared ($JHK_{\rm S}$) photometry of the cluster field, as well as multi-object and long-slit optical spectroscopy for a large
sample of stars in the field. We applied classical photometric analysis, as well as more sophisticated methods using the CHORIZOS code, to determine the reddening to the cluster. We 
then plotted dereddened HR diagrams and determined cluster parameters via isochrone fitting.}    
   {We have identified a large population of luminous supergiants confirmed as cluster members via radial velocity measurements. We find nine red supergiants (plus one other candidate) and two yellow supergiants. We also identify a large population of OB stars. Ten of them are bright enough to be blue supergiants. The cluster lies behind $\approx7.5$~mag of extinction for the preferred value
of $R_{V}=2.9$. Isochrone fitting allows for a narrow range of ages between 12 and 16~Ma. The cluster radial velocity is compatible with distances of $\sim6$ and $\sim10$~kpc. The shorter distance is 
inconsistent with the age range and Galactic structure. The longer distance implies an age $\approx12$~Ma and a location not far from the position where some Galactic models place the far end of the Galactic Bar.}
 {VdBH~222 is a young massive cluster with a likely mass $> 20\,000\:M_{\sun}$. Its population of massive evolved stars is comparable to that of large associations, such as Per OB1. Its location in the inner Galaxy, presumably close to the end of the Galactic bar, adds to the increasing evidence for vigorous star formation in the inner regions of the Milky Way.}

\keywords{open cluster and associations: individual: VdBH~222 - stars: Hertzsprung-Russell (HR) and C-M diagrams, early-type, evolution, supergiant - techniques: photometric and spectroscopic - Galaxy: structure}
\maketitle
%

\section{Introduction}

   The open cluster VdBH~222 was found by \citet{VdBH222} in a two-colour survey of a $\sim12\degr$ wide strip of the southern Milky Way extending from $l\sim250\degr$ to $l\sim360\degr$.  \citet{piatti} present the results of CCD BVI Johnson-Cousins photometry in VdBH~222 and conclude  that it appears to be a young open cluster formed by a vertical main sequence and a conspicuous group of luminous, typically red supergiant stars. They derived a colour excess of $E(V-I)=2.4\pm0.2$, a distance from the Sun of $6.0\pm2.7$ kpc, and an age of $60\pm30$~Ma. They comment that this cluster is among the most reddened and distant open clusters known in the direction towards the Galactic centre.

In the past few years, several clusters of red supergiants have been discovered in a small region of the Milky Way, close to the base of the Scutum-Crux Arm and the tip of the Long Bar, between $l=24\degr$ and $l=29\degr$ \citep{figer2006, davies2007, clark2009, negueruela11, carlos2012}. These clusters are so heavily obscured that, until now, the only members observed are the RSGs. Population synthesis models suggest that the clusters must contain very large stellar populations to harbour so many RSGs \citep{davies2007}. RSGC2 = Stephenson 2 ($l=26\fdg2$, $b=-0\fdg1$) is a strong candidate to be the most massive young cluster in the Galaxy. It is located close to a region where a strong over-density of red supergiants has been found. \citet{negueruela12} studied its possible connection to this over-density. They conclude that RSGC2 is not an isolated cluster, but part of a huge structure most likely containing hundreds of red supergiants, with radial velocities compatible with the terminal velocity at this Galactic longitude and a distance $\sim6$~kpc. 

Previously, \citet{negueruela11} had studied the spatial extent of another cluster of red supergiants, RSGC3 ($l=29\fdg2$, $b=-0\fdg2$), finding that is part of an association that includes smaller clusters of red supergiants. This estimate was later confirmed by the discovery of another cluster of RSGs in its vicinity \citep{carlos2012}. The connection of the RSGC3 association to RSGC2 and the other RSG clusters is unclear. There have been suggestions that this Scutum complex represents a giant star formation region triggered by dynamical excitation by the Galactic bar, whose tip is believed to intersect the Scutum-Crux Arm close to this region \citep{davies2007,garzon1997}.

If this scenario is correct, a similar structure would be expected close to the opposite end of the Galactic Long Bar. 
\citet{carlos} carried out an study of the distribution of red clump giants at longitudes between $l=340\degr$ and $l=352\degr$ with $\mid$ $b$$\mid$$<2\degr$, using the VISTA Variable Survey (VVV) and 2MASS catalogues. They conclude that the observations show the presence of a clear over-density of stars with associated recent stellar formation, which they interpret as the tracers of the Long Bar. They derived an angle for it with the Sun-Galactic Centre line of $41\degr\pm5\degr$, touching the disc near $l=+27\degr$ and $l=348\degr$. Some other studies also support placing the tip of the bar around~$350\degr$ \citep[e.g.][]{babusiaux2005}, although the model of \citet{benjamin2005} assumes a longer and more inclined bar ending around~$340\degr$.

\citet{davies2012} have recently studied a likely massive cluster (Mercer~81) located at the centre of a cavity in a large \ion{H}{ii}-region in the direction of G338.4+0.1, finding the following parameters: a distance of $11\pm2$~kpc , a reddenning $A_{V}=45\pm15$, and an age of $3.7^{+0.4}_{-0.5}$~Ma. With these values, they suggest that it is perhaps placed in the same region of the Galaxy as the far end of the Galactic bar. More recently, \citet{sebas2014} have found a moderate mass (a few $10^3\:M_{\sun}$) for the very young open cluster VVV CL086 ($l=340\fdg0$, $b=-0\fdg3$) located at a similar distance. These locations are roughly compatible with the far tip of the Galactic bar in the model of \citet{benjamin2005}. In the model of \citet{carlos}, they would be placed in the Norma arm tangent. 
     
   In terms of structure of the central regions of the Galaxy, the line of sight towards VdBH~222 is particularly interesting ($l=349\fdg13$; $b=-00\fdg44$), since it intersects the expected location of the far tip of the Galactic bar in the model of \citet{carlos}. Indeed,  the range of ages and distances given by \citet{piatti} allows for just such a placement. Moreover, since it appears to be a very distant object, but only suffers moderate reddening, it will permit the study of the run of extinction in this direction. Here, we investigate this possibility with new photometry and spectroscopy of the cluster. Our data show that VdBH~222 is younger and more distant than suggested by \citet{piatti} and must be located at the far end of the Galactic bar. Its moderate extinction gives it a particular value, since (i) it allows for detection of the blue star population at optical wavelengths, permitting an accurate calibration of the stellar population models that have hereto been employed to determine the properties of the RSG-dominated clusters; and (ii) it provides an excellent laboratory for studying massive star evolution.

In this paper, we identify the evolved stellar population of VdBH~222 via spectroscopy and present photometry from the $U$ to the $K$ band that allows a much better characterisation of the cluster parameters. The paper is organised as follows. In Section~\ref{obs}, we describe the photometric and spectroscopic runs in which data were gathered for this work. In Section~\ref{results}, we analyse the data set and the uncertainties derived from the high reddening, which are used to derive the most likely cluster parameters in Section~\ref{discussion}.

\section{Observations and data}
\label{obs}
In this paper we present a comprehensive set of observations of VdBH~222. The New 
Technology Telescope (NTT) at the La Silla Observatory (Chile) was used to obtain near-infrared (near-IR) photometry with the SOFI instrument. EFOSC2 on the same telescope was used to obtain optical photometry, intermediate-resolution long-slit spectroscopy of the candidate RSGs and low-resolution 
multi-object mask spectra of fainter stars. Finally, the AAOmega fibre-fed multi-object spectrograph, mounted on the Anglo-Australian Telescope (AAT) at the Australian Astronomical Observatory (Siding Springs, Australia), was used to obtain moderately high-resolution spectra of the brightest members around the IR \ion{Ca}{ii} triplet. The observations and reduction procedures are described in the following sections.

\subsection{Optical photometry}
\label{optical}

We obtained $UBVR$ photometry of VdBH~222 using the ESO Faint Object Spectrograph and Camera (EFOSC2) \citep{buzzoni} on the NTT on the night of 24 June 2012. The instrument was equipped with CCD\#40, which is a Loral/Lesser, thinned, AR-coated, UV-flooded, and MPP chip that is controlled by ESO-FIERA. It covers a field of view of $4\farcm1 \times 4\farcm1$ and has a pixel scale of $0\farcs12$.

We obtained several series of different exposure times in each filter to achieve accurate photometry for a broad magnitude range. In the filters $U$ and $B$ we did not take short-time exposures, because we did not expect any cluster member to be saturated. The central position for the cluster and the exposure times are presented in Table~\ref{tab1}.

\begin{table}
\caption{Log of the optical photometric observations taken at the NTT on June 2012 for VdBH~222.\label{tab1}}

\centering
\begin{tabular}{c c c}
\hline
\hline
\noalign{\smallskip}
VdBH~222 & $RA = 17h 18m 48.66s$ & $DEC = -38\,^{\circ} 17\arcmin 32.2\arcsec$ \\
&(J2000)&(J2000)\\
\noalign{\smallskip}
\end{tabular}
\begin{tabular}{c c c c}
\hline
\noalign{\smallskip}
&\multicolumn{3}{c}{Exposure times (s)}\\
Filter & Long times & Intermediate times & Short times \\
\noalign{\smallskip}
\hline
\noalign{\smallskip}
U & 2400 & $-$ & $-$ \\
B & 1800 & 120 & $-$ \\
V & 1200 & 600 & 120;60;20 \\
R & 300 & 150 & 60;10;5 \\
\noalign{\smallskip}
\hline
\end{tabular}
\end{table}

Three standard fields from the list of \citet{landolt}, PG~1633, SA~110A and SA~110B, were observed several times during the night to trace extinction and provide standard stars for the transformation. Their images were processed for bias and flat-fielding corrections with the standard procedures using the CCDPROC package in IRAF\footnote{IRAF is distributed by the National Optical Astronomy Observatories, which
are operated by the Association of Universities for Research in Astronomy, Inc., under cooperative agreement with the National Science Foundation}. Aperture photometry using the PHOT package inside DAOPHOT (IRAF, DAOPHOT) was developed on these fields with
the same aperture (30 pixels) for all filters.

The reduction of the images of VdBH~222 was done with IRAF routines for the bias and flat-field corrections. Photometry was obtained by
point-spread function (PSF) fitting using the DAOPHOT package \citep{stetson1987} provided by IRAF. The apertures used are close to 
the full width at half maximum (FWHM). In this case, we used a value of eight pixels for each image in all filters. To construct the PSF empirically,
we automatically selected bright stars (typically 25 stars). After this,
we reviewed the candidates and discarded those not fulfilling the
requirements for a good PSF star. Once we had a list of PSF
stars ($\approx 20$), we determined an initial PSF by fitting the best
function of the five options offered by the PSF routine inside
DAOPHOT. We allowed the PSF to be variable (of order 2) across the
frame to take into account the systematic pattern of PSF variability
with the position on the chip. We needed to perform an aperture correction of 22 pixels for each frame in all filters. Finally, we obtained the instrumental magnitudes for all stars. Using the standard stars, we carried out the atmospheric extinction
correction and transformed the instrumental magnitudes to the standard system using the PHOTCAL package inside IRAF.

We have optical photometry for 1004 stars in  the field. In the filter $U$ we could not detect the same number of objects as in the other filters. Therefore, the number of stars having the colour $(U-B)$ is only 295. 
In Table 7, we list the identification number of each star, their RA and DEC coordinates in J2000, and the value of $V$,
$(B-V)$, $(U-B)$, and $(V-R)$  with the error and the number of measurements for each magnitude and colour. The number of measurements in $(U-B)$ is always 1 and, therefore, not shown in the Table 7.
The  error  in  each  fitted value is computed by summing in quadrature the contribution to the total error made by each individual  error in the observation file variables. 

\subsection{Near-IR photometry} 

Near-IR $JHK$ images were obtained using the Son OF Isaac (SOFI) near-IR spectrograph and camera \citep{moorwood} on the NTT on 13 July 2008. SOFI was equipped with a Rockwell 1024$\times$1024 HgCdTe Hawaii array, providing a pixel scale of $0\farcs288$ pixel and a field of view of $4\farcm92 \times 4\farcm92$ in its large field (LF) mode. The observations were carried out under photometric conditions. 

These near-IR observations consisted of series of several integrations repeated at dithered positions on the detector. In this case, we have five integrations of 3 s exposure time at seven positions on the CCD randomly determined by the {\tt img\_obs\_Jitter} template. The total effective exposure, resulting from the co-addition of the 35 short jittered exposures, is 105 s in each filter.

The standard reduction of near-IR data includes the dark subtraction and flat-field correction of individual frames
that are median combined to build a sky image. The sky image is subsequently subtracted from the individual
frames that are then shifted and finally stacked into one single image. The reduction was performed using the SOFI pipeline {\tt sofi\_img\_jitter}, which is part of the SOFI subsystem of the VLT data flow system (DFS). The {\tt sofi\_img\_jitter} tool carries out the standard near-IR reduction, including additional steps for cross-talk effect correction and bad-pixel cleaning as described in the SOFI pipeline user manual (VLT-MAN-ESO-19500-4284).

The procedure for obtaining the photometry from the reduced frames was the same as described in Section~\ref{optical} for the optical images of VdBH~222.

The next step is to transform these instrumental magnitudes to the 2MASS magnitude system \citep{Skrutskie2006}. For this, we selected those stars in our field having $JHK_{{\rm S}}$ magnitudes in the 2MASS catalogue with photometric errors smaller than 0.03 magnitudes in every filter (about 70 stars). A linear transformation was carried out between instrumental and 2MASS magnitudes. No colour term was needed, since a simple shift in zero point
results in a good transformation. This was checked by plotting the transformed magnitudes against 2MASS magnitudes for all the stars with 2MASS magnitudes in the field, finding that the best fit corresponds to a straight line of slope 1 in all three filters. 

We have near-IR photometry for 2180 stars. In Table 8 we list the number of each star, their RA and DEC coordinates in J2000, and the value of $J$,
$H$ and $K_{{\rm S}}$ with the photometric error for each magnitude. There is only one measurement for each filter. Only 480 stars have optical photometry  too, and their identification number in Table 7 is shown in the second column.

Astrometric referencing of our images was made using positions of 84 stars from the 2MASS catalogue. The
PSFs of these stars are not corrupted by the CCD over-saturation effects. We used IRAF tasks {\tt ccmap/cctran} for the astrometric
transformation of the image. Formal rms uncertainties of the astrometric fit for our images are $<0\farcs10$ in both right ascension and declination.

\subsection{Long-slit spectroscopy}
Long-slit spectroscopy of ten candidates RSGs was obtained on 26 June 2012, using EFOSC2 in spectroscopic mode.  We used grism \#17, which covers the 6900--8765\AA\ range with a nominal dispersion of 0.9\AA/pixel. The resolution element is 6.5\AA\ with the $1\farcs0$ slit ($R\sim1200$). Since the CCD\#40 is thinned, it presents heavy fringing over most of this spectral range. To correct it, internal halogen flats were taken at the position of each exposure. This procedure reduces the fringing from as much as 10\% to less than 2\%. Data were reduced using the Starlink software packages CCDPACK \citep{Draper2000} and FIGARO \citep{Shortridge1997}. We used standard procedures for bias subtraction and flat-fielding. Wavelength calibration was achieved by using HeAr arc lamp spectra, taken during the afternoon. Because of this, wavelength calibration is not accurate enough to measure radial velocities, even though the rms for the wavelength solution is lower than $\approx0.5$\AA.

\subsection{Mask spectroscopy}

Observations were obtained on 25 June 2012 using EFOSC2 in the multi-object mode. Masks are inserted in the slit wheel with a number
of slitlets punched at the position of the stars to be observed. We used the punch tool with punch \#5, which creates slitlets of width $1\farcs02$
 and length $8\farcs6$, to create two different masks containing 20 candidate blue stars selected from a combination of the photometry of \citet{piatti} and 2MASS data. Each mask was observed with grisms \#3 and \#16. Grism \#3 covers the nominal range 3050--6100\AA\ with a nominal dispersion of 1.5\AA/pixel and a resolution element of 11.6\AA. Grism \#17 covers the nominal range 6015--10320\AA\ with a nominal dispersion of 2.12\AA/pixel and a resolution element of 16\AA. However, the actual spectral range observed depends very strongly on the position of the star in the CCD (the Y distance to the centre of the CCD). The dependence is so strong that two objects situated in the upper third of the image gave no usable spectrum with grism \#16, because the dispersion range had been displaced into the IR. 
The reduction was carried out using standard IRAF procedures. Wavelength calibration was achieved by using HeAr arc lamp spectra, taken during the afternoon. Though internal flats were taken at each position for the grism \#16 exposures, the fringing is only effectively removed for objects close to the centre (in X) of the CCD. 

\subsection{Fibre spectroscopy}

A field centred on VdBH~222 was observed on the nights of 30 August 2010, and 20 July 2011, with the fibre-fed dual-beam AAOmega spectrograph on the 3.9~m Anglo-Australian Telescope (AAT) at the Australian Astronomical Observatory. The instrument was operated with the Two Degree Field ("2dF") multi-object system as front end. Light is collected through an optical fibre with a projected diameter of $2\farcs1$ on the sky and fed into the two arms
via a dichroic beam splitter with crossover at 5\,700\AA. Each arm of the AAOmega system is equipped with a 2k$\times$4k E2V CCD detector (the red arm CCD is a low-fringing type) and an AAO2 CCD controller. Owing to the high reddening to the cluster and short exposure times, only the red arm registered usable spectra. The red arm was equipped with the 1700D grating, blazed at 10\,000\AA. This grating provides a resolving power $R=10\,000$ over
slightly more than 400\AA. The central wavelength was set at 8\,600\AA. The exact wavelength range observed for each spectrum depends on the position of the target in the 2dF field.

Data reduction was performed using the standard automatic reduction pipeline {\tt 2dfdr} as provided by the AAT at the time. Wavelength calibration was achieved with the observation of arc lamp spectra immediately before each target field. The lamps provide a complex spectrum of He+CuAr+FeAr+ThAr+CuNe. The arc line lists were revised, and only those lines actually detected were given as input for {\tt 2dfdr}. This resulted in very good
wavelength solutions, with rms always $<0.1$ pixels.

Sky subtraction was carried out by means of a mean sky spectrum, obtained by averaging the spectra of 30 fibres located at known blank location. The sky lines in each spectrum were evaluated and used to scale the mean sky spectrum before subtraction.

The 2010 observation was a short test run with fibres allocated only to a few bright stars in the cluster regions and some bright targets in its immediate surroundings. Unfortunately, VdBH~222 is too compact to allow observations of a significant number of cluster members in a single configuration. In the 2011 run, targets were selected with a set of criteria similar to those used in \citet{negueruela12}, which were designed to identify red luminous stars. Here we only report on the 11 objects observed in the area covered by our photometry, all of which turn out to be supergiants with similar radial velocities.

\section{Results}
\label{results}
\subsection{Spectroscopy}
\label{spectroscopy}
To summarise, we obtained long-slit spectra with NTT/EFOSC2 of ten red supergiants. These spectra are not affected by fringing because of effective removal. For nine of these RSGs and two yellow supergiants, we also have AAOmega intermediate-resolution spectra in the \ion{Ca}{ii} triplet region that are useful for measuring radial velocities. For the two yellow supergiants and the brightest objects in the photometric sequence, we obtained NNT/EFOSC2 mask spectroscopy with two grisms. Beyond 7500\AA, these spectra are affected by fringing. Since the mask observations cannot be done at parallactic angle, only the yellow supergiants have usable spectra shortwards of 4700\AA\ .

\begin{figure*}[ht]
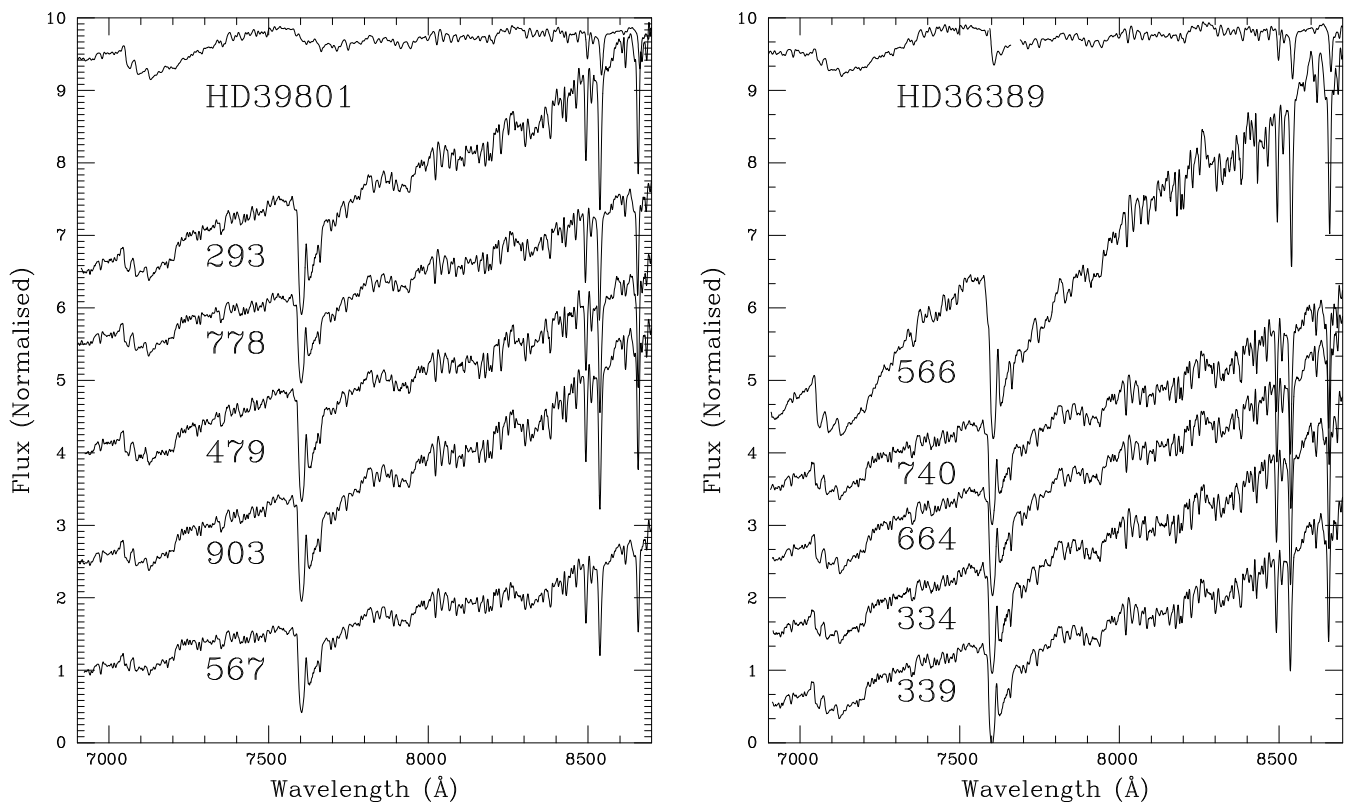

\resizebox{\columnwidth}{!}{\includegraphics[clip]{efosc1.eps}}
\resizebox{\columnwidth}{!}{\includegraphics[clip]{efosc2.eps}}
\caption{NTT/EFOSC2 spectra for ten red supergiant members observed in the cluster. The comparison spectra are from the MILES library \citep{sanchez2006}. They have had all the telluric features removed. Since they are not subject to strong interstellar reddening, re-scaling to unity at 7700\AA\ conveys a very different appearance to that of the cluster stars. \label{rsgs_efosc}}
\end{figure*}

\subsubsection{Red supergiants}

Figure~\ref{rsgs_efosc} shows the NTT/EFOSC2 long-slit spectra of ten red supergiant candidates, selected from the 2MASS CMD, in the photographic IR. Given the complexity of late-type spectra and the impossibility of removing the effects of interstellar extinction, the spectra were simply re-scaled to unity at 7700\AA. With the exception of 566, all the stars have spectral types close to M0. The main classification criteria in this range are the triple TiO bandhead from the $\gamma$ system at 7055, 7088, 7126\AA\ \citep[which blends into an atmospheric water vapour band at 7186\AA\ and becomes visible at K5;][]{turnshek85}, the triple TiO bandhead of the $\delta$ system at 8432--42--52\AA\ \citep[noticeable from M1; see][for a discussion]{negueruela12}, and the bandhead system at 8250\AA, which becomes observable at M0, but is blended into the weak atmospheric water-vapour system at 8227\AA. Comparison to standards would suggest that 567 is the earliest object, with a spectral type between K5 and M0; 293, 479, 778, and 903 are not far from M0; 334, 339, 664, and 740 are close to M1; and 566 is the only object decidedly later than M2. 

\begin{figure}
\resizebox{\columnwidth}{!}{\includegraphics[clip]{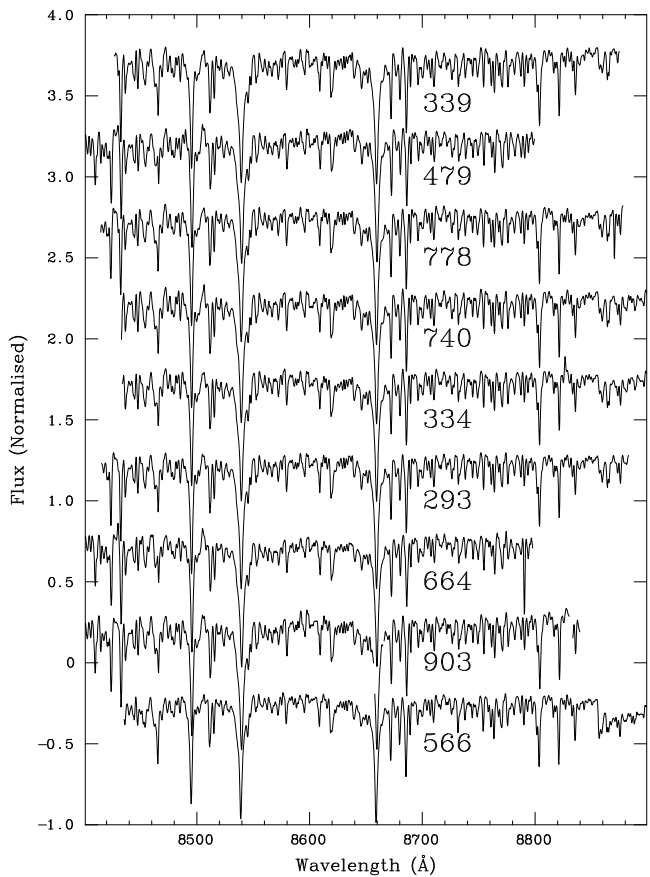}}
\caption{AAOmega spectra for nine red supergiant members observed in the cluster.\label{omega}}
\end{figure}

Figure~\ref{omega} shows the AAOmega higher resolution spectra for nine of the stars. Unfortunately, 567 could not be observed. Classification criteria in this region at similar resolution are discussed in \citet{negueruela12}.  The ratio of the 8514\AA\ feature to the neighbouring \ion{Ti}{i}~8518\AA\ ($>>1$ in all cases) confirms that all these objects are supergiants. The best calibrated indicators are the equivalent width (EW) of the two strongest lines in the \ion{Ca}{ii} triplet and the \ion{Ti}{i} dominated blend at~8468\AA. Based on them, almost all the objects are likely to have luminosity class Iab, with the exception of 740, which looks more luminous. Since the 8432--42--52\AA\ feature is hardly noticeable (and not visible in all spectra), we can resort to the TiO $\delta$(0,0) $R$-branch bandhead at 8860\AA, where available. Based on this, 740 and 334 are M1; 293 is M0.5; and 339 and 479 are M0. Again 566 is clearly later with a spectral type close to M3. For all the other spectra, we cannot be more precise than a classification between K7 and M2. The only obvious discrepancy between the two datasets comes from 339, which appears earlier here than in the NTT spectra taken one year later. Even though spectral variability is not unusual in red supergiants \citep[e.g.][]{levesque2007}, further monitoring of this object is needed before variability can be claimed. A summary of the spectral types derived is given in the upper panel of Table~\ref{tab:rsgs}.

\begin{table}
\centering
\caption{Spectral types and other parameters for luminous supergiants.\label{tab:rsgs}}
\begin{tabular}{lccc}
\hline
\hline
\noalign{\smallskip}
ID &$i^{a}$ & Spectral&$v_{{\rm LSR}}$\\
 && type&(km\,s$^{-1}$)\\
\noalign{\smallskip}
\hline
\noalign{\smallskip}
293 &12.2&M0.5\,Iab&$-$98$^{c}$\\
334 &11.9&M1\,Iab&$-$91\\
339 &11.4&M0--1\,Iab&$-$93\\
479 &11.5&M0.5\,Iab&$-$96\\
566 &11.3&M3.5\,Iab&$-$106\\
567 &11.9&K5--M0\,I& $-$\\
664 &11.9&M1\,Iab&$-$100\\
740 &11.5&M1\,Ia&$-$97$^{c}$\\
778 &11.4$^{b}$&M0\,Iab&$-$97\\
903 &12.1&M0.5\,Iab&$-$103\\
\noalign{\smallskip}
\hline
 \noalign{\smallskip}
371 &10.6&G0\,Ia$^{+}$&$-$100\\
505 &11.3&G2\,Ia&$-$102\\
 \noalign{\smallskip}
\hline
\end{tabular}
\begin{list}{}{}
\item[]$^{a}$ $i$ magnitudes from DENIS.
\item[]$^{b}$ The supergiant is blended with a foreground star (774 in our photometry) in DENIS and 2MASS.
\item[]$^{c}$ The value is an average of two measurements.
\end{list}
\end{table}

The AAOmega spectra can also be used to measure radial velocities. The procedure used, based on cross-correlation with a grid of  of model spectra from the POLLUX database \citep{palacios10}, is described in \citet{negueruela12}. The radial velocities measured were transformed into the local standard of rest (LSR) reference system using IRAF {\it rvcorrect} package, using the standard solar motion ($+20\:{\rm km}\,{\rm s}^{-1}$ towards $l=56\degr$ , $b=23\degr$).
The average of all measurements is $v_{{\rm LSR}}=-99\:{\rm km}\,{\rm s}^{-1}$, with a standard deviation of $4\:{\rm km}\,{\rm s}^{-1}$. All observed values are within 2$\sigma$ of the mean, a value perfectly consistent with the expected intrinsic variability of red supergiants \citep{negueruela12}. Two stars (293 and 740) were observed in 2010 and 2011, and they present differences of $-5.6\pm2\:{\rm km}\,{\rm s}^{-1}$ and  $-6.5\pm2\:{\rm km}\,{\rm s}^{-1}$. We looked for a possible systematic difference between the 2010 and 2011 observations, but it is not obvious. The 2010 data give $v_{{\rm LSR}}=-97\pm5\:{\rm km}\,{\rm s}^{-1}$ for seven measurements, and the 2011 data give $v_{{\rm LSR}}=-100\pm2\:{\rm km}\,{\rm s}^{-1}$ for six datapoints. 

\begin{figure}
\resizebox{\columnwidth}{!}{\includegraphics[clip,angle=-90]{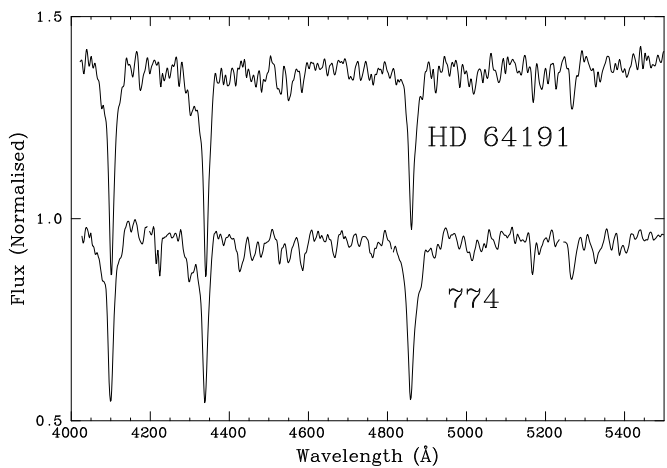}}
\caption{Spectrum of star 774, which dominates the blue magnitudes for the red supergiant 778, compared to the early F-type giant HD~64191, taken from the library of \citet{jacoby}.\label{fgiant}}
\end{figure}

Though 778 looks completely normal in the spectra shown, its blue spectrum is dominated by another star, 774, which is blended with it in the photometry of \citet{piatti}, and also in 2MASS. Figure~\ref{fgiant} shows the blue spectrum of 774 compared to an early F-type giant. Given the very low reddening, it is clear that 774 is a foreground F0--2\,IV star and not a cluster member.

\subsubsection{Yellow supergiants}

\begin{figure}
\resizebox{\columnwidth}{!}{\includegraphics[clip, angle=-90]{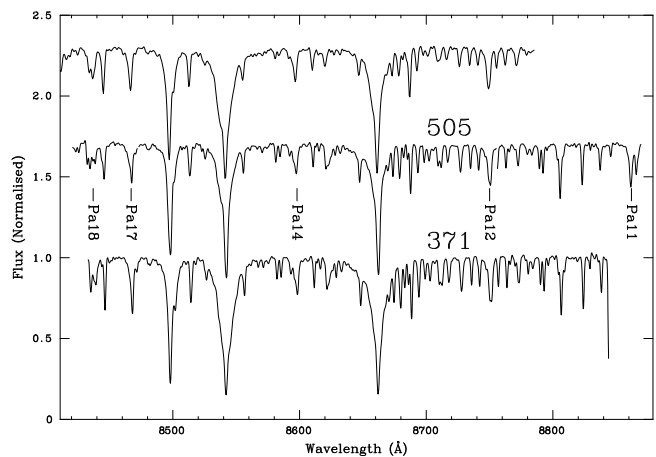}}
\caption{AAOmega spectra of the two yellow supergiants in VdBH~222. The comparison spectrum (top) is the G0\,Ia MK standard HD\,18391. Note that Pa~13, 15 and 16 are blended with the three lines of the \ion{Ca}{ii} triplet. The \ion{Fe}{i}~8621\AA\ line is blended with an interstellar band in the cluster stars.\label{yellow}}
\end{figure}

The AAOmega spectra of two yellow supergiants in VdBH~222 are shown in Fig.~\ref{yellow}, compared to the spectrum of the G0\,Ia MK standard HD\,13891 from the atlas by \citet{andrillat95}. In this temperature range, the intensity of the Paschen lines declines strongly with spectral type, but also with increasing luminosity class. Star 505 has Paschen lines slightly weaker than HD\,13891, but a stronger metallic spectrum. This indicates a later spectral type, in the G1-2\,Ia range. Star 371, on the other hand, presents stronger Paschen lines than the standard, but also a much stronger metallic spectrum. This combination only makes sense if it is a more luminous star. Indeed the EW of the \ion{Ca}{ii} triplet in 371 is rather higher than in any of the other two ($>35$\AA). We conclude that it is a G0\,Ia$^{+}$ extreme supergiant, similar to the very luminous F-type stars found in the starburst cluster Westerlund~1 \citep{clark10}.

\begin{figure}
\resizebox{\columnwidth}{!}{\includegraphics[clip,angle=-90]{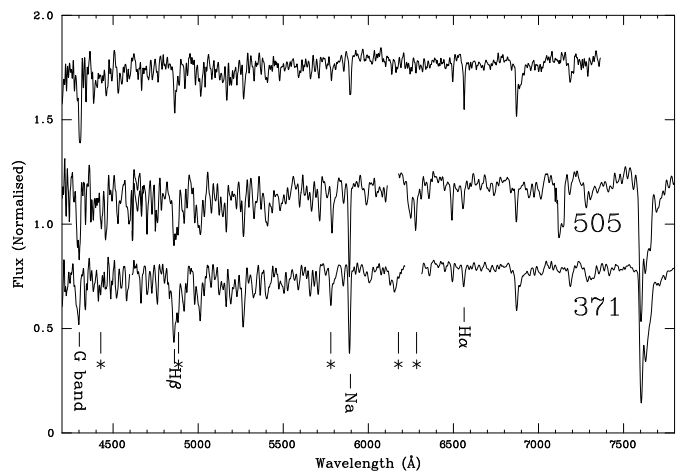}}
\caption{NTT/EFOSC2 spectra of the yellow supergiants in VdBH~222, compared to the G3\,Ib supergiant HD\,191010, from the atlas of \citet{jacoby}. The \ion{Na}{i} doublet in the cluster stars includes an important interstellar component. Other prominent interstellar features are marked with ``*''. The spectrum of 505 presents a strong reduction artefact around 7100\AA.\label{hblues}}
\end{figure}

The low-resolution spectra of the two stars are displayed in Fig.~\ref{hblues}. The spectra are a combination of the grism \#3 and grism \#16 NTT/EFOSC2 spectra. As a comparison, the low-resolution spectrum of the G3\,Ib supergiant HD\,191010 from the atlas by \citet{jacoby} is shown. No supergiants of higher luminosity and similar spectral type are included in the atlas. If the ratio of the G-band to H$\gamma$ is taken as an indicator of the spectral type, HD\,191010 is clearly later than the two cluster stars. The stronger metallic spectrum demonstrate the higher luminosity. H$\beta$ is much stronger in the cluster stars, suggesting an strong contribution from the diffuse interstellar band (DIB) at 4885\AA. Other prominent DIBs are also indicated in Fig.~\ref{hblues}.

The spectral types derived and radial velocities measured for the two yellow supergiants are shown in the bottom panel of Table~\ref{tab:rsgs}.

\begin{figure}
\resizebox{\columnwidth}{!}{\includegraphics[clip]{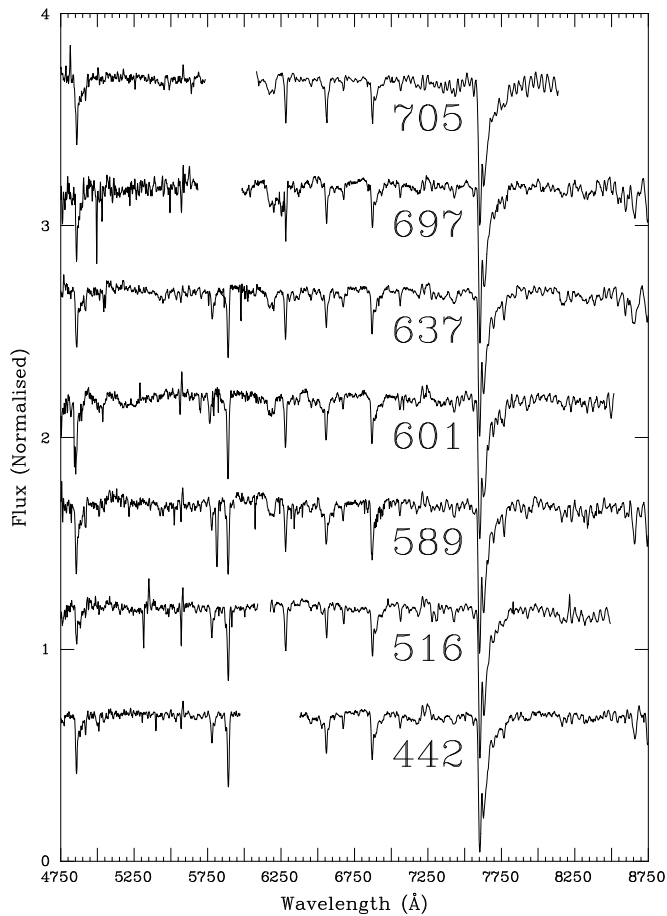}}
\caption{NTT/EFOSC2 spectra of the brightest ``blue'' members of VdBH~222. The gap marks the shift from grism \# 3 (blue) to grism \# 16 (red). There is a sky-subtraction artefact around 7240\AA\ in most spectra. \label{blues}}
\end{figure}

\subsubsection{Brightest blue stars}

Unfortunately, the spectra of stars just above the cluster main-sequence are quite noisy. The atmospheric window around 8500\AA, which is useful for spectral classification, cannot be used because of heavy fringing.  The spectra shown in Fig.~\ref{blues} are dominated by interstellar and atmospheric features. A few objects (for example, 589 or 442) seem to display the strong narrow Paschen lines typical of high-luminosity B-type stars, but they are difficult to separate from fringing. H$\alpha$ is present in all the spectra, but always affected by sky subtraction issues. All the spectra clearly display the \ion{He}{i}~6678 and~7065\AA\ lines. The interstellar \ion{Na}{i} doublet is blended in a single feature, together with the \ion{He}{i}~5875\AA\ line. The prominent DIBs at 5780 and 5797\AA\ are also blended into a single feature in the spectra where that region is covered by grism \#3. The DIB at 6284\AA\ is also clearly seen in most spectra. H$\beta$ appears as a strong feature in all spectra, but it must be blended with the strong DIB at 4885\AA. The spectra with higher S/N in the blue show the \ion{He}{i}~4922\AA\ line partially blended into this feature. The \ion{He}{i}~5016\AA\ line is affected by a strong reduction artefact in most spectra, but may be seen in 442 and 589.

The detactability of many \ion{He}{i} lines at this low resolution indicates 
that all the stars are B-type objects, either supergiants or early-B ($<$B5) main-sequence or giant stars. If the Paschen lines in 442 and 589 are true, they support the supergiant classification. As H$\alpha$ is in absorption in all the spectra, they are unlikely to be high-luminosity (class Ia) supergiants, but rather objects of luminosity class Ib or II \citep[cf.][]{negueruela10}.

\subsection{Photometry}

\subsubsection{Observational HR diagram}
\label{HR}

\begin{figure}
\resizebox{\columnwidth}{!}{\includegraphics[clip]{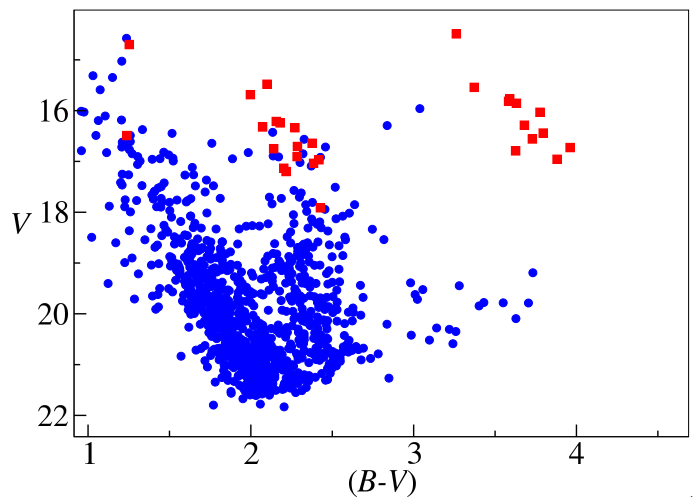}.}
\caption{$V/(B-V)$ diagram for all stars in the field. Red squares represent stars spectroscopically observed.\label{VBV}}
\end{figure}

\begin{figure}
\resizebox{\columnwidth}{!}{\includegraphics[clip]{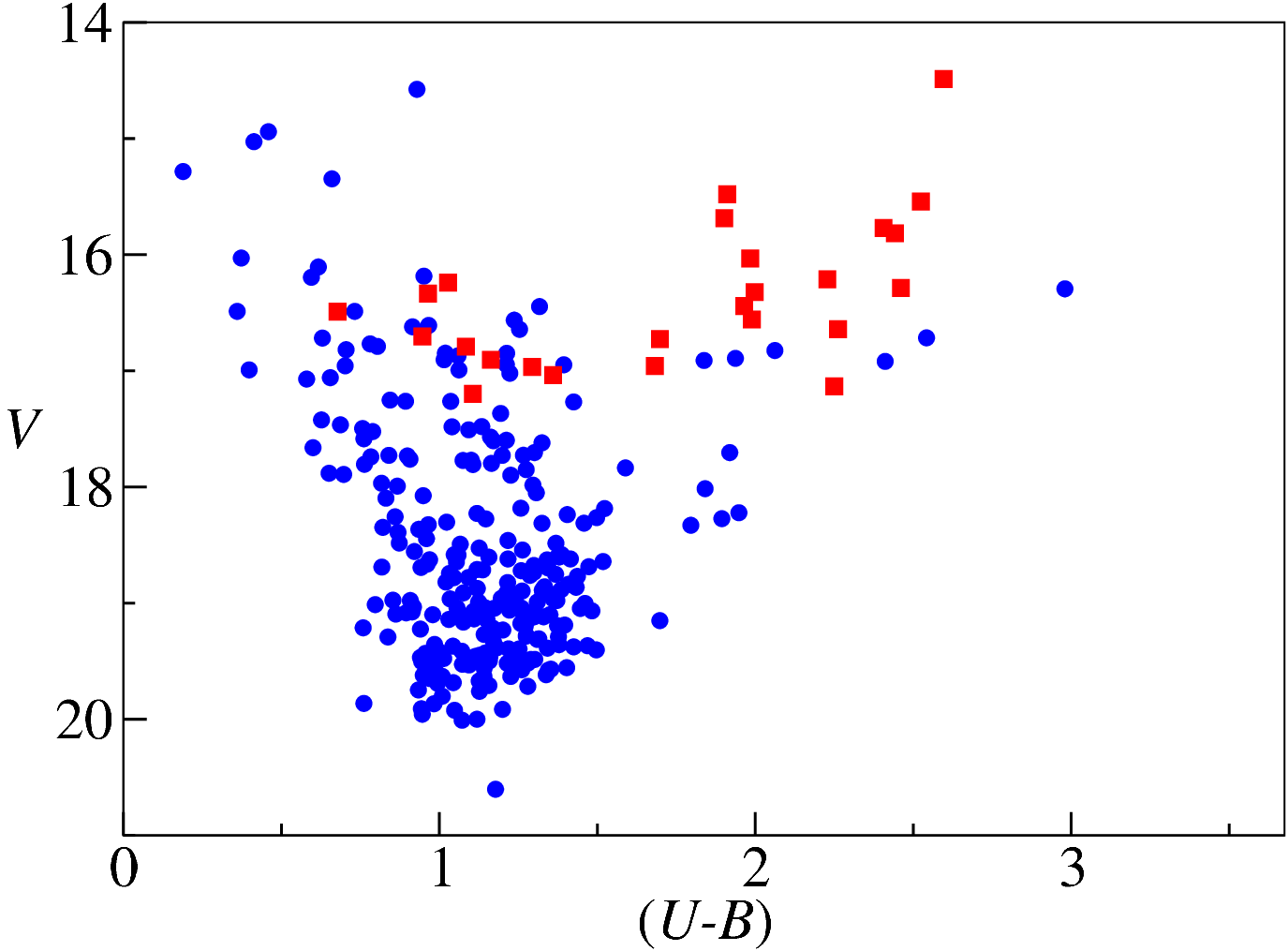}.}
\caption{$V/(U-B)$ diagram for all stars in the field. Red squares represent stars spectroscopically observed.\label{VUB}}
\end{figure}

\begin{figure}
\resizebox{\columnwidth}{!}{\includegraphics[clip]{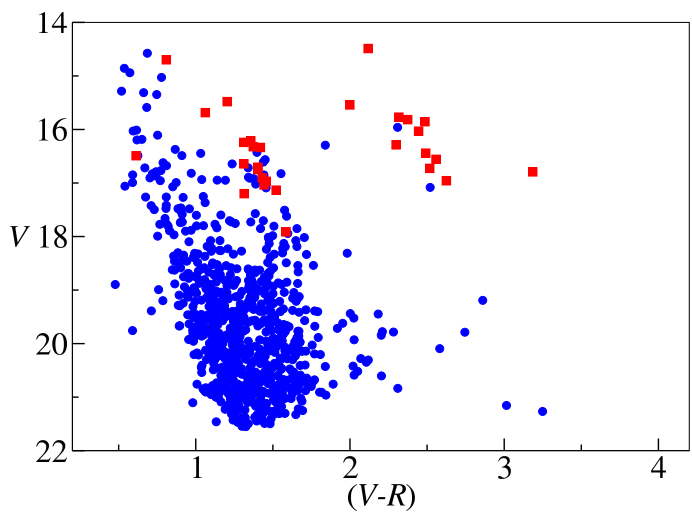}}
\caption{$V/(V-R)$ diagram for all stars in the field. Red squares represent stars spectroscopically observed.\label{VVR}}
\end{figure}

\begin{figure}
\resizebox{\columnwidth}{!}{\includegraphics[clip]{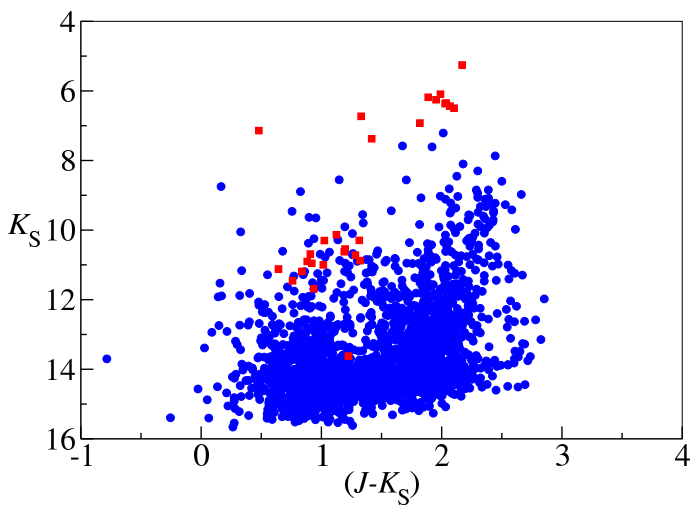}}
\caption{$K_{{\rm S}}/(J-K_{{\rm S}})$ diagram for all stars in the field. Red squares represent stars spectroscopically observed.\label{KJK}}
\end{figure}

We started the photometric analysis by plotting the $V/(B-V)$, $V/(U-B)$, $V/(V-R)$ and $K_{{\rm S}}/(J-K_{{\rm S}})$ diagrams for all stars in our field (see Figures~\ref{VBV},~\ref{VUB},~\ref{VVR} and~\ref{KJK}). The confirmed OB stars fall on top of the vertical sequence in the middle of all diagrams, while two unreddened stars belong to the sequence to the left of all diagrams. The red supergiants are the clump of bright stars to the right of all diagrams. The first step is to remove those objects that clearly belong to the left sequence (until $V\sim 20$) in both $V/(B-V)$ and $V/(V-R)$ diagrams, corresponding to the foreground population. Generally, if they have any measurement in $(U-B)$, their positions in the $V/(U-B)$ diagram are incorrect, too.

The second step is to find the main sequence of early-type stars that correspond to the brightest blue stars that are spectroscopically observed. For this purpose, we used the near-IR photometry. We calculated the IR reddening-free parameter $Q$=$(J-H)-1.8(H-K_{{\rm S}})$, and we selected stars with a value of $Q$$\leq$$0.08$. This cut is expected to only select early-type stars. From the selection, we chose those objects whose $(J-K_{{\rm S}})$ values are compatible with those of members spectroscopically observed ($0.5$$\leq$$(J-K_{{\rm S}})$$\leq$$1.5$; see Figure~\ref{KJK}). 

These colour cuts leave us with an initial list of likely members of the cluster, all of which have measurements in both optical and near-IR bands. To be as exhaustive as possible, we did not reject stars lacking near-IR photometry, but having the complete set of $UBVR$ magnitudes, since reliable astrophysical information can be obtained from their optical colours. These stars were included in the optical analysis. We obtain the parameters of the cluster, analysing the two diagrams: $M_{V}/(B-V)_{0}$ and $M_{K_{{\rm S}}}/(J-K_{{\rm S}})_{0}$.

\subsubsection{Determination of the $R_{5495}$  value and a preliminary $E(B-V)$}
\label{jesus} 

We used the latest version of the software package CHORIZOS \citep{jesus2004} 
to process the $UBVJHK_{{\rm S}}$ photometry for the nine brightest OB
star candidates. We used the stellar grid presented by \citet{jesus2013a}
 fixing the luminosity class to either 5.0 (main sequence) or 1.0
(supergiants), setting distance as a dummy parameter (i.e. using colours rather
than magnitudes as input), and leaving three free parameters: effective
temperature ($T_{\rm eff}$), amount of monochromatic colour excess
($E(4405-5495)$), and type of extinction ($R_{5495}$). We used the family of
extinction laws recently derived by \citet{jesus2013b}, which have
been shown to produce significantly better results for stars with $A_V \sim 6$ than
those of \citet{cardelli1989}. The monochromatic values $E(4405-5495)$ and $R_{5495}$ are used, because, unlike their band-integrated counterparts, $E(B-V)$ and $R_V$, they provide a description of the extinction law \citep{jesus2013b}

In those cases where the photometric uncertainties were smaller than 0.04
magnitudes, they were set to that value for the CHORIZOS analysis. There are
three reasons to take this step:

\begin{enumerate}
 \item Zero-point calibration. As shown by \citet{jesus2005, jesus2006, jesus2007}, there are systematic uncertainties of the order of 0.02-0.03 magnitudes in the comparison between Johnson $UBV$ magnitudes and spectrophotometry (either observed or theoretical) due to calibration issues.
 \item Highly reddened objects. The above uncertainties become even greater in
the case of objects with anomalous colours caused by high extinction. In those cases, red leaks in the filters can produce undesired offsets in the measured magnitudes.
 \item The accuracy of the extinction law outside the range of values for which
it was calculated \citep[see][for details]{jesus2013b}. The inaccuracy affects not only the optical range but, for extinctions with $A_V \gtrsim 6$, can also affect near-IR photometry. Both \citet{cardelli1989} and \citet{jesus2013b} use the \citet{rieke1985} law in the near-IR, so systematic effects may appear when the slope adopts different values
there and the extinction is high enough.
\end{enumerate}

Increasing the photometric uncertainties provides a way of reducing the
systematic effects and providing accurate results at the cost of making them
too conservative.

Results are shown in Table~\ref{tabCHO} for the reduced $\chi^2$ of each best fit, the three fitted parameters, and three additional derived quantities ($E(B-V)$, $A_V$, and $(m_V-A_V)$). The main conclusions that can be derived follow:

\begin{enumerate}
 \item The results using main-sequence or supergiant models are very similar.
This is expected, given that  gravity induces only small changes in the
optical/near-IR colours for late-O/early-B stars (with the only exception of supergiants with extreme mass loss). For this reason, only one set of results is shown.
 \item The values of $\chi^2_{\rm red}$ are reasonable for seven of the nine
stars, indicating that those are ``good fits''. For the other two stars, 705 and 520, the values of $\chi^2_{\rm red}$ are too high. We have checked that this is not due to a large IR excess in the $K$ band. Inspection of the images suggests that the near-IR magnitudes for 520 are contaminated by another star. This does not seem to be the case for 705.
 \item As expected when using such large uncertainties, precise values of
$T_{\rm eff}$ cannot be obtained. All we can say from the CHORIZOS results is that the stars are either of spectral type O or early B. Star 627 appears to be cooler than the rest. 
 \item $R_{5495}$ is close to 2.8 for eight stars, the only exception being star
601, which has a higher value that is closer to the standard.
 \item The monochromatic colour excesses show only small variations around 2.70.
The band-integrated values are lower by $\sim$0.1 magnitudes, as expected for this amount and this type of extinction. 
 \item The values of $A_V$ cluster around 7.5 magnitudes. The largest deviation
corresponds to star 601, the object with the largest $R_{5495}$.
 \item As usual in this type of analysis, $E(4405-5495)$ and $R_{5495}$ are
anticorrelated. Therefore, $A_V$ shows relatively small uncertainties compared to either one of them. Other correlations between quantities lower the uncertainty in $(m_V-A_V)$ with respect to that of $A_V$.
\end{enumerate}

\begin{table*}
\centering
\caption{Parameters obtained for the nine brightest stars of the cluster using CHORIZOS code. \label{tabCHO}}
\begin{tabular}{lrr@{$\pm$}lr@{$\pm$}lr@{$\pm$}lr@{$\pm$}lr@{$\pm$}lr@{$\pm$}l}
\hline
\noalign{\smallskip}
Name & \multicolumn{1}{c}{$\chi^2_{\rm red}$} & \multicolumn{2}{c}{$T_{\rm
eff}$} (kK) & \multicolumn{2}{c}{$R_{5495}$} & 
\multicolumn{2}{c}{$E(4405-5495)$} & \multicolumn{2}{c}{$E(B-V)$} &
\multicolumn{2}{c}{$A_V$} & \multicolumn{2}{c}{$m_V-A_V$} \\
\noalign{\smallskip}
\hline
\noalign{\smallskip}
442 &  0.31 & 37.6&7.5 & 2.78&0.05 &           2.55&0.05 & 2.46&0.05 &
7.04&0.09 & 9.19&0.07 \\
 516 &  1.93 & 41.8&5.5 & 2.90&0.05 & $\;\;\;\;$2.60&0.03 & 2.51&0.03 &
7.48&0.06 & 8.86&0.04 \\
520$^{a}$ & 10.00 & 33.3&8.6 & 2.85&0.06 &           2.72&0.07 & 2.61&0.07 &
7.68&0.11 & 8.88&0.11 \\
589 &  1.12 & 36.5&7.8 & 2.75&0.05 &           2.82&0.05 & 2.70&0.05 &
7.68&0.09 & 9.28&0.08 \\
601 &  2.72 & 38.3&7.1 & 3.27&0.06 &           2.56&0.04 & 2.48&0.04 &
8.31&0.08 & 8.44&0.06 \\
627 &  1.46 & 24.3&5.4 & 2.97&0.07 &           2.60&0.08 & 2.50&0.08 &
7.66&0.13 & 9.76&0.13 \\
637 &  3.15 & 33.6&4.5 & 2.90&0.05 &           2.69&0.04 & 2.59&0.04 &
7.75&0.08 & 9.63&0.07 \\
697 &  0.12 & 35.4&8.1 & 2.79&0.05 &           2.60&0.06 & 2.51&0.05 &
7.19&0.10 & 9.55&0.09 \\
705 &  4.79 & 37.8&7.6 & 2.68&0.05 &           2.87&0.05 & 2.74&0.05 &
7.60&0.09 & 9.38&0.08 \\
 \noalign{\smallskip}
\hline
\end{tabular}
\begin{list}{}{}
\item[]$^{a}$ Inspection of the IR images suggests that the star is blended with another object so its $JHK_{{\rm S}}$ magnitudes are suspect.
\end{list}
\end{table*}

\subsubsection{Determination of reddening and photometrical spectral types}

Alternatively, the colour excess $E(B-V)$ for cluster members may be derived from the photometry, using standard calibrations. This procedure is only accurate when the values of $(U-B)$ are known. Therefore we restricted our initial analysis to stars with $U$ magnitudes.

 The $Q$ parameter is a measure of the colour of an O or B star, corrected for the effect of interstellar reddening \citep{johnson1953}. The $Q$ parameter is defined as

\begin{equation}
\label{Qdefinition}
 Q=(U-B)-\frac{E(U-B)}{E(B-V)}(B-V)
\, .
\end{equation} 

Observations and analyses have shown that there is a change in the slope of the reddening line as a function of spectral type \citep{johnson1958}. The expression for this change is

\begin{equation}
\label{slope}
\frac{E(U-B)}{E(B-V)}=X+0.05E(B-V)
\, ,
\end{equation} 
where $X$ depends weakly on spectral type (or, correspondingly, intrinsic colour).

VdBH~222 has a large population of early M-type red supergiants (Section~\ref{spectroscopy}), indicating that its age must be between 10~Ma and 25~Ma \citep{mermilliod1981}. In this age range, blue members around the main-sequence turn-off are expected to have spectral types around B1--B2. For these spectral types, the value of $X$ in Eq.~\ref{slope} is $0.69$ and the intrinsic colour of stars around the turn-off is then $(B-V)_{0}$ $\approx-0.2$ \citep{fitzgerald}. We can use this hypothesis to give a first estimate of $E(B-V)$, using the $(B-V)$ colour of stars in the upper main sequence. The estimated value of $E(B-V)$  is $\approx2.5$. This value is fully compatible with the result obtained by CHORIZOS for the nine brightest likely members of the cluster. In Table~\ref{tabCHO} we can see that the median value for $E(B-V)$ is $2.56$ (excluding 520) and therefore,
 
\begin{equation}
\frac{E(U-B)}{E(B-V)}=0.82
\, .
\end{equation} 

We then calculated the $Q$ parameter for all likely cluster members with $(U-B)$ values. Using $Q$,  we assigned photometric spectral types \citep{johnson1953} to all these stars. From the spectral types, we calculated individual values of the colour excess $E(B-V)$, using the intrinsic colours from \citet{johnson1958} and 

 \begin{equation}
E(B-V)=(B-V)-(B-V)_{0}
\, .
\end{equation}

In Table 4 we display the values of $Q$, photometric spectral types, $(B-V)_{0}$, and $E(B-V)$ for likely members with $(U-B)$ colours. The brightest blue stars in the cluster have $V$ magnitudes comparable of those of M supergiants. These objects must therefore be supergiants so the calibration for main-sequence stars cannot be used \citep{johnson1958}. We made a rough division of the blue stars and used calibrations for supergiants for objects brighter than $V=17$, calibrations for giants 
for objects with $17\leq V \leq18.5$, and assumed that fainter objects are on the main sequence.

We obtained an average value of $E(B-V)=2.45\pm0.15$ using only those stars believed to be on the main sequence. Inclusion of the stars taken as giants or blue supergiants does not change this average. We therefore take this as the average colour excess for the cluster. Consequently, all the stars in the cluster blue sequence for which we lack $U$ magnitudes have been dereddened with this average value. In Table 5 we show the value of $(B-V)_{0}$ for these stars.

There is a small difference between this value and the one obtained in the CHORIZOS analysis ($2.56\pm0.09$ when star 520 is excluded). We note the high amount the differential reddening and the fact that the CHORIZOS value is based on only eight stars, but direct comparison of the values obtained with the two methods for the eight stars in Table~\ref{tabCHO} shows that the difference is likely to be systematic, even though the two values are compatible within their respective errors. The reasons for this difference are connected to those listed in Sect.~\ref{jesus} and likely stem from the fact that the standard relations of \citet{johnson1953} are derived for stars that are much less reddened than cluster members.

We then estimated the extinction to each star using its individual colour excess when available or the cluster average, and the total-to-selective extinction ratio derived in Section~\ref{jesus}. The average $R_{5495}$ is 2.83 for all stars except star 601. If we use $A_{V}=R_{V}\cdot E(B-V)$, this corresponds to an average $R_{V}=2.90$. We calculated intrinsic magnitudes as

 \begin{equation}
V_{0}=V-2.90\cdot E(B-V)
\, .
\end{equation}  

In Tables 4 and 5 we display the value of $V_{0}$ for all likely members of the cluster. In Figure~\ref{Fig1} we display the spatial distribution of likely members. The values of $(B-V)_{0}$ and $V_{0}$ are used to build the $M_{V}/(B-V)_{0}$ diagram and to determine the distance and age of the cluster. In Table 4, we can see that our rough division of the stars into main sequence, giants, and supergiants may not have always been accurate. At least two stars classified as giants (627 and 705) have intrinsic magnitudes as bright as the stars classified as supergiants. This is due to differential reddening, but has little effect on the intrinsic parameters derived, since the intrinsic colours of B-type giants and supergiants at a given spectral type differ by a few hundredths of a magnitude at most.

The red supergiants have been dereddened following the procedure of \citet{fernie}, which transforms the observed $E(B-V)$ to an equivalent $E(B-V)$ for early type stars to account for colour effects. Star 566 has not been included in the $M_{V}/(B-V)_{0}$ diagram because its $(U-B)$ colour is 
very different from those of the other supergiants, suggesting that its magnitudes are contaminated by a blue star.

Since we also have near-IR photometry for most likely cluster members, we may also employ the $M_{K_{{\rm S}}}/(J-K_{{\rm S}})_{0}$ diagram to help us determine the distance and age of the cluster. For this purpose, we began by using those likely members with photometric spectral types and near-IR photometry values that can be used to calculate their individual colour excesses $E(J-K_{{\rm S}})$. Using the intrinsic colours of \citet{winkler1997}, we could calculate $E(J-K_{{\rm S}})$ for each star (49 stars in total). The average of $E(J-K_{{\rm S}})$ is $1.07\pm0.14$. This value, however, gives rise to two difficulties:

\begin{enumerate}

\item For a standard reddening law, $E(J-K_{{\rm S}})$ can be converted to $E(B-V)$ using a calibration relation from \citet{fiorucci2003}. For hot stars, the relation is approximately (estimated for $T_{{\rm eff}}=17800K$):
\begin{equation}\label{exceso}
E(J-K_{{\rm S}})=0.476\cdot E(B-V)+0.007\cdot E(B-V) \cdot E(B-V)\,.
\end{equation} 

When we replace our value $E(B-V)=2.45\pm0.15$ in Eq.~\ref{exceso}, we obtain $E(J-K_{{\rm S}})\sim1.20$, which is somewhat higher ($0.13$~mag) than the value obtained from the individual excesses. The $E(B-V)=2.56$ obtained from the CHORIZOS analysis would imply an even higher $E(J-K_{{\rm S}})$ for the standard law. This may again be related to a non-standard extinction law.

\item However, a direct comparison of the intrinsic colours of \citet{winkler1997} with the \citet{marigo2008} isochrones shows that the models predict $(J-K_{{\rm S}})_{0}$ values $\approx-0.06$ bluer than the empirical calibration. This discrepancy might be due to the use of $K$-band magnitudes by \citet{winkler1997} instead of the $K_{{\rm S}}$ employed by \citet{marigo2008}, but we note that other widely used empirical calibrations of IR colours show the same effect, with the calibration of \citet{wegner94} resulting in an even greater discrepancy. 
\end{enumerate}

In view of this, we were forced to choose a colour excess that fits the isochrone rather than using the intrinsic colours. The value required to fit the isochrone is $E(J-K_{{\rm S}})=1.16$.  

Once more, we find small, but not negligible, discrepancies that reflect inconsistencies between the different calibrations of intrinsic colours and uncertainties in the shape of the extinction law that become important when we study clusters with high reddening. As discussed later, \citet{carlos} find evidence of an IR extinction law in this direction that is slightly more transparent than the standard \citet{rieke1985} law. In spite of this, we built the $M_{K_{{\rm S}}}/(J-K_{{\rm S}})_{0}$ diagram using the standard $A_{K_{{\rm S}}}=0.67E(J-K_{{\rm S}})$, since we can expect deviations from the standard law to be small at these moderate values of $E(J-K_{{\rm S}})$.

\subsubsection{Determination of distance and age}
\label{distance}

\begin{figure}
\resizebox{\columnwidth}{!}{\includegraphics[clip]{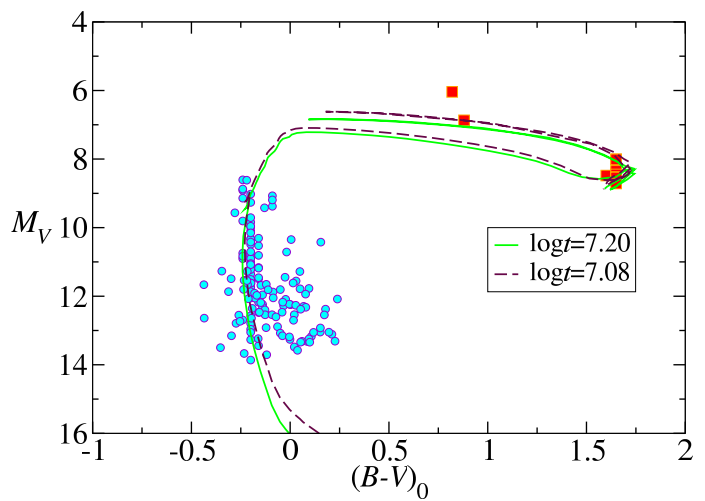}}
\caption{$M_{V}/(B-V)_{0}$ diagram for selected likely members. Red squares represent red and yellow supergiants spectroscopically observed in the field. The solid line is the isochrone of \citet{marigo2008} corresponding to $\log t = 7.20$, unreddened and displaced to $DM=13.8$. The dashed line is the isochrone of \citet{marigo2008} corresponding to $\log t = 7.08$, unreddened and displaced to $DM=14.5$. The value used for $R$ is the standard value of $3.1$. \label{fig10}}
\end{figure}

\begin{figure}
\resizebox{\columnwidth}{!}{\includegraphics[clip]{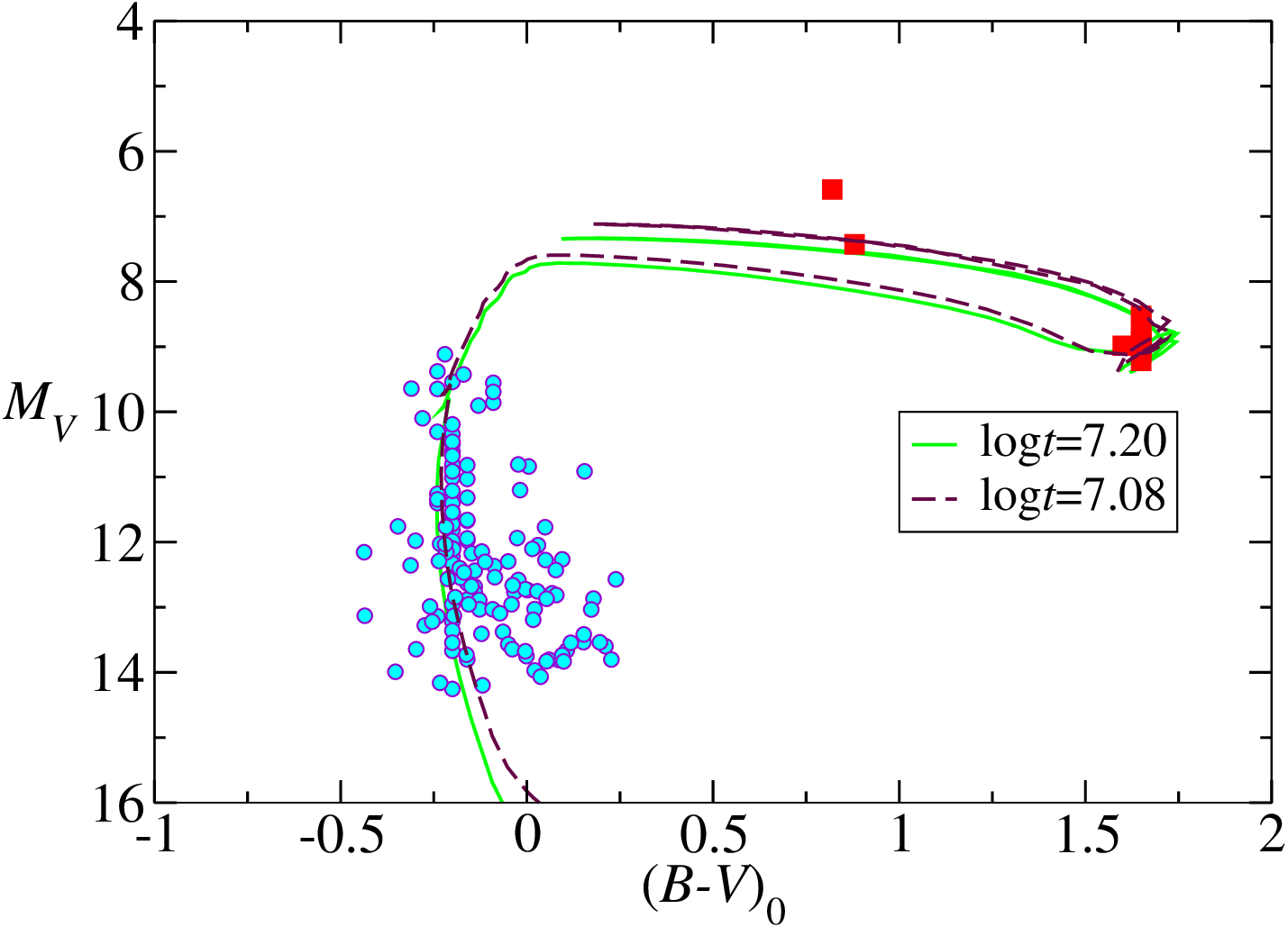}}
\caption{$M_{V}/(B-V)_{0}$ diagram for selected likely members. Red squares represent red and yellow supergiants spectroscopically observed in the field. The solid line is the isochrone of \citet{marigo2008} corresponding to $\log t = 7.20$, unreddened and displaced to $DM=14.3$. The dashed line is the isochrone of \citet{marigo2008} corresponding to $\log t = 7.08$, unreddened and displaced to $DM=15.0$. The value used for $R_{V}$ is $2.9$. \label{fig11}}
\end{figure}

Finally, we used the intrinsic colours in optical and near-IR bands to build the $M_{V}/(B-V)_{0}$ and $M_{K_{{\rm S}}}/(J-K_{{\rm S}})_{0}$ diagrams and to determine the distance and age to the cluster by eye-fitting isochrones. In Figures~\ref{fig10} and~\ref{fig11} we plot the $M_{V}/(B-V)_{0}$ diagram for the selected likely members of the cluster, calculating $A_{V}$ as $R\cdot E(B-V)$ with values of $3.1$ and $2.9$ for $R_{V}$, respectively. In Fig.~\ref{fig10}, we plot two isochrones from \citet{marigo2008} corresponding to $\log t=7.20$, unreddened and displaced to $DM=13.8$ ($d=5.8$~kpc), and $\log t=7.08$, unreddened and displaced to $DM=14.5$ ($d=7.9$~kpc). In Fig.~\ref{fig11} the same two isochrones must be shifted to different distances, because of the lower $A_{V}$. The $\log t=7.20$ isochrone is displaced to $DM=14.3$ ($d=7.2$~kpc), and the $\log t=7.08$ isochrone is displaced to $DM=15.0$ ($d=10.0$~kpc). We can observe that the two isochrones fit well both the main sequence and the position of the red supergiants, because there is a small degeneracy between distance, age, and reddening that can be broken with the near-IR data; however, older or younger isochrones fail to fit the turn-off and the position of the red supergiants at the same time. The figures clearly illustrate the enormous impact of the value of $R$ chosen on the distance determined at such high values of $E(B-V)$. 

In Figure~\ref{fig12}, we display the $M_{K_{{\rm S}}}/(J-K_{{\rm S}})_{0}$ diagram. In this figure, we plot two isochrones from \citet{marigo2008} that correspond to $\log t=7.20$, unreddened and displaced to $DM=14.3$ ($d=7.2$~kpc), and $\log t=7.08$, unreddened and displaced to $DM=14.9$ ($d=9.5$~kpc). If we consider that the typical uncertainty in the by-eye fit is $\pm0.2$ mag in $DM$, we see that the values obtained from the optical and IR HR diagrams using $R_{V}=3.1$ are not consistent. On the other hand, when we use  $R_{V}=2.9$, the isochrones fit the data at the same distance for both datasets. Therefore, we have a range of possible solutions going from a 16~Ma cluster at $\approx7$~kpc to a 12~Ma cluster at $\approx10$~kpc.

\begin{figure}
\resizebox{\columnwidth}{!}{\includegraphics[clip]{graficaMk_JK0.eps}}
\caption{$M_{K_{{\rm S}}}/(J-K_{{\rm S}})_{0}$ diagram for selected likely members. Red squares represent red and yellow supergiants spectroscopically observed in the field. The solid line is the isochrone of \citet{marigo2008} corresponding to $\log t = 7.20$, unreddened and displaced to $DM=14.3$. The dashed line is the isochrone of \citet{marigo2008} corresponding to $\log t = 7.08$, unreddened and displaced to $DM=14.9$. \label{fig12}}
\end{figure}


\begin{figure*}[ht]
\resizebox{12 cm}{!}{\includegraphics[angle=-90]{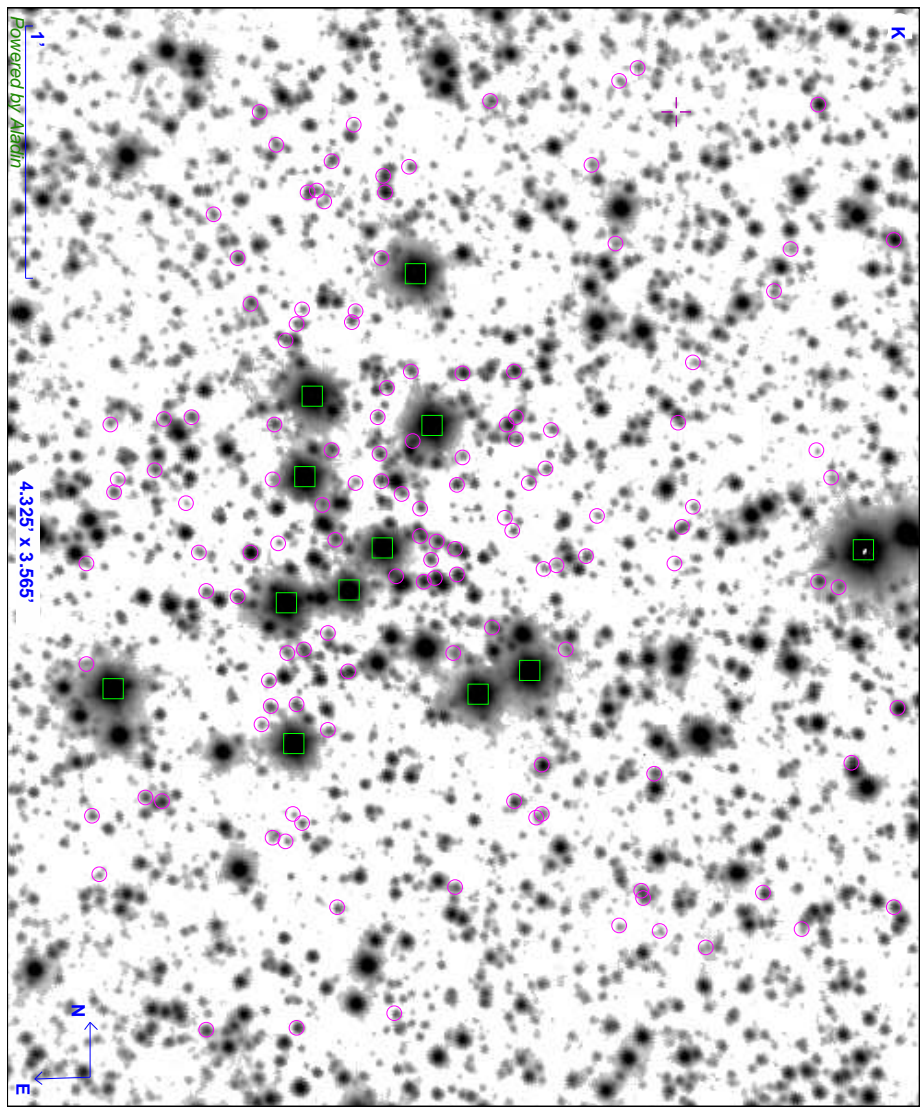}}
\centering
\caption{Finding chart for VdBH~222. The image is our $K$-band frame. Stars marked with a rectangle are yellow or red supergiants. Stars inside a circle are other likely members. North is up and East is left. \label{Fig1}}
\end{figure*}

\subsubsection{Theoretical HR diagram for red supergiants}
\label{HRtheo}

 In obscured clusters, the distance is the most uncertain parameter and must be estimated from the radial velocity  \citep[e.g.][]{davies2007} or the run of extinction \citep[e.g.][]{negueruela11}. In many cases, cluster parameters have been derived exclusively from the RSG population \citep[e.g.][]{davies2007,clark2009,negueruela10}. We can repeat this procedure here as a further check for consistency.

In Table 6, we present the 2MASS photometry for eleven red and yellow SGs in VdBH~222. The errors in determining of some magnitudes are high because of saturation problems, but we found no other broad-band near-IR photometry for these stars in the literature. The $K_{\rm S}$ magnitude for 778 is not available, so it has been excluded from the analysis. We derived the excesses in $(J-K_{{\rm S}})$ for each star, taking the average intrinsic colours for each spectral type from \citet{carlos2012} and \citet{koornneef1983}. Assuming standard reddening, $A_{K_{{\rm S}}}=0.67E(J-K_{{\rm S}})$, we calculated $K_{{\rm S}}$ for each object. Then, we calculated $M_{K_{{\rm S}}}$ for each RSG, using a  distance modulus for the cluster $DM=15.0$, which is typical of the values found of the analysis of $UBVJHK_{{\rm S}}$ photometry.

\setcounter{table}{5}

\begin{table*}
\caption{Photometric data for red supergiants in VdBH~222 from 2MASS and our photometry.}
\centering
\begin{tabular}{lcccccccccc}
\hline\hline
Star&$J$&$E_{J}$&$H$&$E_{H}$&$K_{{\rm S}}$&$E_{K_{{\rm S}}}$&$E(J-K_{{\rm S}})$&$M_{K_{S}}$&$\log (L/L_{\sun})^{1}$&$\log (L/L_{\sun})^{2}$\\ 
\hline\hline
293&8.600&0.024&7.164&0.021&6.497&0.021&1.073&$-$9.2&4.5&4.6\\
334&8.394&0.021&6.993&0.024&6.365&0.023	&	0.979&$-$9.3&4.5&4.6\\
339&8.070&0.020&6.735&0.026&6.182&0.051	&	0.858&$-$9.4&4.6&4.7\\
371&7.681&0.020&6.732&0.020&6.223&0.026	&	1.048&$-$9.5&5.0&$-$	\\
479&8.207&0.023&6.846&0.018&6.254&0.021	&	0.913&$-$9.4&4.6&4.6	\\
505&8.793&0.020&7.880&0.020&7.375&0.027	&	0.978&$-$8.3&4.6&$-$	\\
566&7.425&0.023&5.971&0.031&5.255&0.018&1.03&$-$10.4&4.9&5.1\\
567$^{3}$&8.742&0.005&7.470&0.016&6.923&0.024	&	0.839&$-$8.6&4.3&4.4\\
664&8.384&0.026&6.974&0.021&6.348&0.020	&	0.976&$-$9.3&4.5&4.6	\\
740&8.084&0.024&6.717&0.020&6.094&0.020	&	0.940&$-$9.5&4.6&4.7	\\
903&8.506&0.026&7.098&0.020&6.437&0.018	&	1.029&$-$9.3&4.5&4.6	\\
\noalign{\smallskip}
\hline
\end{tabular}
\begin{list}{}{}
\item[]$^{1}$ Using the calibrations of \citet{levesque2005}.
\item[]$^{2}$ Using the calibration of \citet{davies2013}, which is not valid for yellow supergiants.
\item[]$^{3}$ 567 is the only star for which $JHK_{{\rm S}}$ is from our photometry. All the other stars have at least one filter saturated, and their magnitudes are from 2MASS.
\end{list}	
\end{table*}

Finally, we used the following expressions taken from \citet{levesque2005} to calculate the intrinsic stellar parameters for the theoretical H-R diagram: 

\begin{equation}
(J-K)_{0}=3.10-0.547(T_{{\rm eff}}/1000\:{\rm K})
\end{equation}

\begin{equation}
 {\rm BC}_{K}=5.574-0.7589(T_{{\rm eff}}/1000\:{\rm K})
\end{equation}

\begin{equation}
\label{leves}
(L/L_{\sun})=10^{-(M_{{\rm bol}}-4.74)/2.5}\, .
\end{equation}

The determined parameters are also listed in Table 6. Recently, \citet{davies2013} have cast doubt on these temperature scale and bolometric corrections, after analysing the spectral energy distributions of a sample of RSGs in the Magellanic Clouds and finding that effective temperatures show little dependence on spectral type. They derive an empirical method to calculate bolometric luminosities from single-band photometry. We used the $M_{K_{S}}$ calculated above to estimate luminosities according to their method. We find values $\sim 0.1$~dex higher than using Eq.~\ref{leves} (but entirely compatible within errors) for all stars.

\begin{figure}[ht!]
\resizebox{\columnwidth}{!}{\includegraphics[clip]{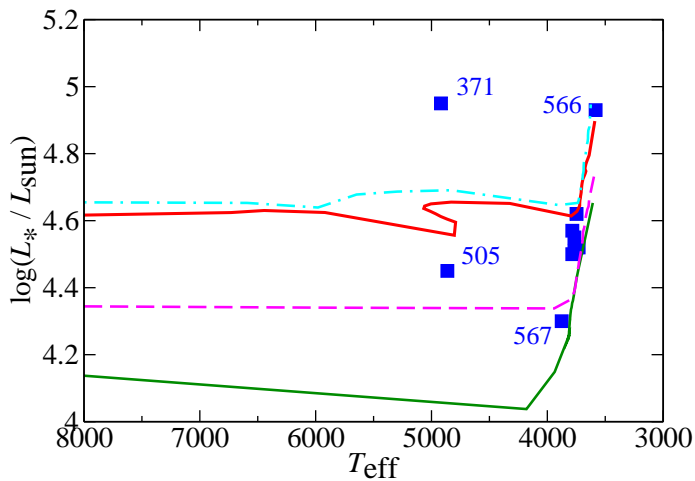}}
   \caption{HR diagram showing the locations of 11 red and yellow supergiants in the cluster, with their positions derived from the spectral classification, assuming a distance to the cluster $d=10$~kpc ($DM=15.0$). We also plot isochrones from \citet{Ekstrom2012}. The dotted lines are the $\log t=7.20$ (16 Ma; top, cyan), and $\log t=7.30$ (20 Ma; bottom, pink) isochrones with high initial rotation. The solid lines are the $\log t=7.08$ (12 Ma; top, red), and $\log t=7.20$ (16 Ma; bottom, green) isochrones without rotation. The size of the symbols represents the errors in  $\log L_{*}$ that are due to observational uncertainties and calibration problems. The errors in $T_{{\rm eff}}$, though difficult to quantify, should be smaller than the symbol size.\label{Figteorica}}
    \end{figure}

In Figure~\ref{Figteorica}, we display the theoretical Hertzsprung-Russell (HR) diagram showing the positions of the eleven supergiants compared to different isochrones from \citet{Ekstrom2012}. The positions of the RSGs are consistent with the high initial rotation isochrone for $18-19$~Ma and an age $\approx13$~Ma for low rotation. Therefore the isochrones without rotation agree fairly well with the Padova isochrones. However, the high-rotation isochrones require a much older age. In addition, we note that this age is not directly comparable to those obtained in previous works \citep[e.g.][]{davies2007,clark2009,marco2013} that use the isochrones of \citet{meynet2000}. With those older isochrones, the age of VdBH~222 at $DM=15.0$ would be around 15 Ma. If we take into account that the recipes adopted for the mass-loss rate during the RSG phase have a very strong impact on their location in the theoretical HR diagram \citep{georgy2012}, to the point that they can loop back bluewards and become yellow hypergiants for the highest mass-loss rates, we have to conclude that an absolute age derivation is too dependent on the isochrones used. We therefore stick to the nominal 
values obtained from the non-rotating isochrones.  

Two stars have unexpected positions in this diagram. Star 567 appears much fainter than all the other stars. This is the only RSG for which we lack intermediate-resolution spectra, hence a measurement of its radial velocity. Its membership must thus remain suspect until a measurement of radial velocity can be obtained. The other outlier is the yellow supergiant 371, which appears much brighter than the other yellow supergiant, 505, in accordance with its extreme spectroscopic appearance.  

\section{Discussion}
\label{discussion}
We have found a large population of yellow and red supergiants in the open cluster VdBH~222. Comparable clusters found towards the base of the Scutum arm have estimated ages in the range 12\,--\,20~Ma, though these values are mostly based on rough determinations of their kinematic distances \citep[e.g.][]{davies2007,davies2008}. The possibility of observing blue stars in VdBH~222 in the $UBV$ bands offers an excellent opportunity to provide much more accurate parameters for this cluster, though we still face the difficulty of the unknown shape of the extinction law in a cluster with $A_{V}$ approaching 8~mag (noting that the same uncertainties afflict analysis of the clusters at the base of the Scutum arm, and other highly reddened clusters). Without a good knowledge of the extinction law, the cluster age and cluster distance cannot be determined with accuracy, because the number of free parameters is so high that the problem becomes degenerate. The analysis of the brightest stars with CHORIZOS strongly suggests a non-standard extinction law, with a value for $R_{V}=2.9$. When this value is used, the optical and IR data give consistent solutions. The isochrone fits allow a range of ages between 12 and 16~Ma (according to the isochrones of \citet{marigo2008}). This range of ages corresponds to a range of distances between 7 and 10~kpc. We can make use of our knowledge of
Galactic structure to choose the more likely solution.

\begin{figure}[ht]
\resizebox{\columnwidth}{!}{\includegraphics[angle=90,clip]{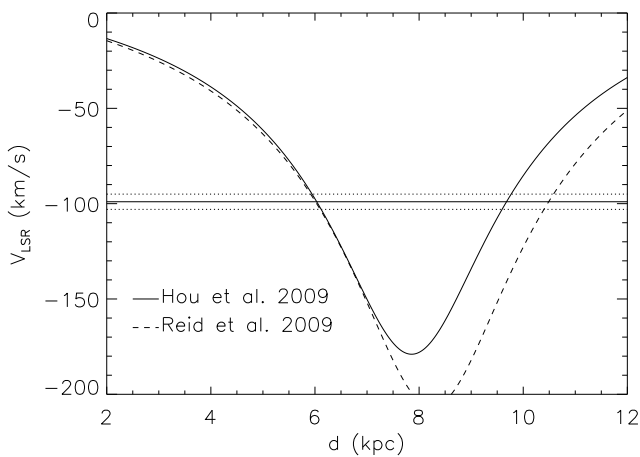}}
\centering
\caption{Variation of radial velocity with respect to the local standard of rest (LSR) due to Galactic rotation, as a function of distance along the line of sight to VdBH~222. The Galactic rotation curve of \citet{brandblitz} has been recalibrated after \citet{hou2009} and \citet{reid2009}. The solid line shows the measured velocity for VdBH~222, while the dashed lines show the 1-$\sigma$ standard deviation.\label{rotcurve}} 
\end{figure}

\subsection{Cluster parameters}

At a distance of 7~kpc (compatible with an age of 16~Ma), adopting a distance from the Sun to the Galactic Centre of 8.3~kpc \citep[e.g.][]{reid2014}, VdBH~222 would be at a galactocentric distance $R_{{\rm G}}=1.9$~kpc. This seems very unlikely to be physically possible, since it would place the cluster too close to the Galactic Centre, in the region where the Galactic bar potential would prevent the formation of any stable structure.

The cluster radial velocity has been measured to be  $v_{{\rm LSR}}=-99\pm4\:{\rm km}\,{\rm s}^{-1}$. Figure~\ref{rotcurve} shows the distribution in radial velocity along the line of sight to VdBH~222, using the Galactic rotation curve of \citet{brandblitz}, recalibrated after \citet{hou2009} and \citet{reid2009}. As can be seen, two values are compatible with the observed velocity. The short distance is $\approx6$~kpc, while the long distance depends on the rotation curve adopted, with acceptable values between $\approx9.5$ and 10.5~kpc.

A distance of 6~kpc ($DM=13.9$) is marginally consistent with the photometric data. The fit to the isochrones, which is worse than for the favoured age range 12--16~Ma, requires an age $\approx 18$~Ma. At this distance, VdBH~222 would be at a galactocentric distance $R_{{\rm G}}=2.7$~kpc. This would place the cluster within or beyond the 3~kpc arm \citep{green2009,sanna2014}, which has a similar radial velocity. However, the maps of neutral \ion{H}{i} by \citet{burton1983} seem to show that the 3~kpc arm lies above the Galactic Plane for $l<352\degr$. Moreover, extinction maps by \citet{carlos} show that extinction does not increase along this line of sight beyond $d=4$~kpc (the position of the Norma arm), suggesting that the 3~kpc arm is very tenuous and unlikely to be associated with vigorous star formation. Moreover, the extinction maps show a very low value of extinction out of 13~kpc in the box centred on $l=349\fdg5$, $b=-0\fdg5$ \citep{carlos}, corresponding to a hole in dark clouds that is perfectly visible in 2MASS wide-field images. In addition, \citet{carlos} find evidence of a more transparent extinction in the near IR, with a slope closer to that of \citet{nishiyama} than to the standard \citet{rieke1985} law. No other spiral arm can be found at this distance, so this value of the distance is not favoured by Galactic structure. A shorter distance, placing the star in the Norma arm, can be safely ruled out. On the one hand, the expected radial velocity would be much lower, around  $v_{{\rm LSR}}=-40\:{\rm km}\,{\rm s}^{-1}$. On the other hand, this would require the cluster to be somewhat older than 20~Ma, in disagreement with the presence of a large population of M-type supergiants.

The other distance permitted by the rotation curve is $\approx10$~kpc, which is compatible with the photometric analysis for a cluster age of $\approx12$~Ma. At this distance, VdBH~222 would also be placed at $R_{{\rm G}}\approx2.4$~kpc, but this time on the far side of the Galaxy. According to \citet{carlos}, this position should be fairly close to the far tip of the Galactic bar. Of course, we can always suspect that the proximity of the Galactic bar could produce strong deviations from the Galactic rotation curve, as demonstrated by objects close to the near end of the bar \citep{reid2014}. On the other hand, the requirement to keep the cluster on this side of the bar prevents a much greater distance anyway. \citet{carlos} find a very strong increase in extinction at a distance $\sim13$~kpc, forcing VdBH~222 to be closer than this distance. Given the uncertainties involved in the isochrone fits and the exact shape of the extinction law, an error in our distance modulus determination of $\pm0.3$ is perfectly acceptable. This narrow range in distance modulus covers distances between 8.7 and 11.5~kpc. The need to keep the cluster outside the 3~kpc ring and on the near side of the bar favours a distance between 10 and 11~kpc, depending on the exact location of the tip of the bar. In this most likely scenario,
VdBH~222 would be located close to the base of the Perseus arm and would represent the analogue of the Scutum Complex clusters. As will be demonstrated in a future work, VdBH~222 is surrounded by a large number of red supergiants with similar radial velocities, supporting its connection with a major star-forming region. Critically, an accurate determination of the spectral types of the blue stars close to the main-sequence turn-off would allow us to fix the extinction law and the exact age and distance at the same time. 

\subsection{Previous work}

Using only $V$ and $I$-band photometry and assuming a standard reddening law, \citet{piatti} estimated an age of $60\pm30$~Ma and a distance $6\pm3$~kpc. These values are roughly consistent with our analysis for a standard extinction law, since the optical data alone can be fit with a cluster age 18~Ma at a distance of 6~kpc. A direct comparison to their photometry is not possible, because we lack the $I$ band. However, taking into account that $E(V-I) = 1.6 \cdot E(B-V)$ for the extinction law of \citet{rieke1985}, their $E(V-I)=2.4$ seems too low when compared to our $E(B-V)=2.45$, even if the extinction law is more transparent than the standard. In the even more heavily reddened open cluster Westerlund~1, the relationship was found to be $E(V-I) \approx1.3 \cdot E(B-V)$ \citep{negueruela2010a}. 

\subsection{Stellar content}
We have identified ten red supergiant and two yellow supergiant members of VdBH~222, as well as around ten OB members that seem bright enough to be blue supergiants. Such a huge population of evolved stars is unprecedented in optically accessible Milky Way clusters. For instance, the open cluster NGC~7419, with a mass $\ga 5\,000\:M_{\sun}$, contains only five red and two blue supergiants \citep{marco2013}. The evolved population of VdBH~222 is comparable to that of the core region of the Per~OB1 association, containing the open clusters NGC~869 and NGC~884 (h and $\chi$ Persei), estimated to have a mass of at least $20\,000\:M_{\sun}$ at an age $\sim14$~Ma \citep{currie2010}. Indeed, simulations of stellar populations by \citet{clark2009b} suggest that a cluster of 10\,--\,12~Ma must have a mass in excess of $30\,000\:M_{\sun}$ to sport 12 cool supergiants. Our optical photometry, which is certainly incomplete, includes $\approx135$ blue stars with dereddened magnitudes in the range $9<V_{0}<14$, all likely to be B5 or earlier. This number results in a lower limit for the cluster mass  $\sim10\,000\:M_{\sun}$ for an age around 12~Ma.
VdBH~222 is thus likely to be comparable in age and mass to RSGC1 \citep{davies2008}, one of the massive clusters found close to the base of the Scutum arm.

Two other clusters found towards the Scutum arm contain a higher number of RSGs, Stephenson~2 \citep{davies2007,negueruela12} and RSGC3 \citep{clark2009}. Both are much more obscured than VdBH~222, making them invisible in the $U$ and $B$ bands, and both are thought to be somewhat older. In spite of the higher number of RSGs, none of the two clusters hosts any yellow supergiant, comparable to the two G-type objects found in VdBH~222. Indeed, this kind of luminous G-type supergiant is very rare in young clusters and only seen in clusters younger than $\sim12$~Ma \citep{clark2013}, lending support to our final conclusion about the age of the cluster. 

The position of star 371 with respect to all the other stars is peculiar. It is the brightest cluster member in all diagrams and looks too bright for the cluster isochrone in Fig.~\ref{Figteorica}, even though its radial velocity confirms it as a cluster member. Its spectrum in the region of the \ion{Ca}{ii} triplet is extreme, looking much more luminous than the G0\,Ia standard HD~18391. However, its actual bolometric magnitude is far from extreme. As calculated in Section~\ref{HRtheo} for $DM=15.0$, it is $M_{{\rm bol}}\approx-7.6$. This is far from the luminosities of the yellow hypergiants thought to be post-RSGs, such as HR~8752 or $\rho$~Cas, and even less than the estimate for the moderately luminous object IRAS~1835$-$06 \citep{clark2014}. Moreover, the very strong absorption spectrum contrasts with the emission-line dominated appearance of the latter. Therefore we speculate that star 371 is more likely to be the descendant of a blue straggler caught in a short evolutionary phase rather than a post-RSG star. Monitoring of this peculiar object to search for variability would undoubtedly cast light on its nature, as highlighted by the recent detection of a binary companion to the 
yellow hypergiant HR~5171 \citep{chesneau2014}.

\section{Conclusions}

We have identified a large population of evolved massive stars in the open cluster VdBH~222, comprising blue, yellow, and red supergiants. The radial velocities of
eleven members have been measured, giving a cluster average $v_{{\rm LSR}}=-99\pm4\:{\rm km}\,{\rm s}^{-1}$. With this radial velocity, VdBH~222 must be a very distant cluster. The near kinematic distance of $\approx6$~kpc is inconsistent with the Galactic structure, since it would place the cluster on the 3~kpc arm in a location where no absorption that could indicate the presence of star formation is found. Therefore we favour the far kinematic distance of $\approx10$~kpc that places the cluster close to the far end of the Galactic bar. Photometric analysis indicates heavy differential reddening with $A_{V}$ values in the $\approx7$--8~mag range.
Lacking accurate spectral types for the blue stars, a fit to the observed magnitudes with the CHORIZOS code cannot constrain the values of $R$ and $E(B-V)$ very strongly, but suggests a value for $R_{V}=2.9$. The optical and near-IR photometric analysis are consistent with a cluster age $\approx12$~Ma at a distance of $\approx10.0$~kpc. 

The cluster is extremely compact with very few members lying outside a radius of $1\farcm5$ from the cluster centre (see Figure~\ref{Fig1}). The observed population of red and yellow supergiants (9+2 confirmed members with one further candidate RSG) indicates that VdBH~222 is a young massive cluster. This is confirmed by the presence of about ten blue stars with $V$ magnitudes comparable to those of the RSGs, which must be blue supergiants. VdBH~222 is thus almost certainly the most massive cluster observable in the $U$ and $B$ bands in the Milky Way. As such, it represents an ideal laboratory for studying the extinction law in the inner Milky Way and the evolution of high mass stars towards the RSG stage.  

The location of VdBH~222 close to the far end of the Galactic bar adds to the evidence for vigorous star formation close to its tip, analogous to the near end. In this respect, we must stress that VdBH~222 is much closer to the Galactic Centre (and so to the nominal tip of the bar) than other massive clusters recently reported in the area \citep{davies2012, sebas2014}. In a future work, we will report on an exploration of the area surrounding VdBH~222 to locate evidence of coeval star formation.

\begin{acknowledgements}

We are very grateful to Mart\'{\i}n Guerrero for providing the reduced SOFI images from which the IR photometry was extracted. The images were obtained
by Mart\'{\i}n Guerrero and Luis F. Miranda, and reduced by Gerardo Ramos-Larios, all of whom are warmly thanked. We thank the referee, Prof. Rolf Kudritzki, for useful comments.
This research is partially supported by the Spanish Ministerio de
Ciencia e Innovaci\'on (MICINN) under
grant AYA2012-39364-C02-02.
JMA acknowledges support for this work by the Spanish Government Ministerio de
Ciencia e Innovaci\'on through grants AYA 2010-17631 and AYA 2010-15081 and by
the Junta de Andaluc\'{\i}a grant P08-TIC-4075.
The AAT observations have been supported by the OPTICON project (observing proposals 2010B/01 and 2011A/014), which is funded by the European Commission under the Seventh Framework Programme (FP7).
This research made use of the Simbad, Vizier, and Aladin services developed at the Centre de 
Donn\'ees Astronomiques de Strasbourg, France and of the WEBDA database, operated at the
Institute for Astronomy of the University of Vienna. This publication
makes use of data products from the Two Micron All Sky Survey, which is a joint project of the University of
Massachusetts and the Infrared Processing and Analysis
Center/California Institute of Technology, funded by the National
Aeronautics and Space Administration and the National Science
Foundation.

\end{acknowledgements}


\end{document}